\documentclass[11pt,twoside]{article}

\usepackage{bm}
\usepackage{latexsym}
\usepackage{amssymb}

\usepackage{amsmath}
\usepackage{amsthm}

\usepackage[all]{xy}
\usepackage{dcolumn}
\usepackage{mathrsfs}

\usepackage[numbers,square,sort&compress]{natbib} 

\usepackage{dsfont}

 \usepackage{amsmath,amsthm,amssymb}
 \usepackage{makeidx,epsfig,lscape}
 \usepackage{color,colortbl}
 \usepackage{fancyhdr}
 \usepackage{xcolor,pict2e}

 \renewcommand{\headrulewidth}{0pt}
 \renewcommand{\footrulewidth}{0.5pt}

 \definecolor{myaqua}{rgb}{0.0,0.5,0.55}
 \definecolor{lightaqua}{rgb}{0.75,0.95,0.95}

 \usepackage[colorlinks = true,
            linkcolor = myaqua,
            urlcolor  = blue,
            citecolor = myaqua]{hyperref}

\usepackage{caption}
\usepackage{floatrow}
\def\lin#1#2{\textcolor[rgb]{0.6,0.6,0.6}{\vspace*{#1mm} \hrule
   height 3 pt \vspace*{#2mm}}}
\def\bt{\begin{tabular}}
\def\et{\end{tabular}}
\def\and{\mbox{ and }}

\def\1{{\bf 1}}

 \def\sectionn#1{\refstepcounter{section}{\color{myaqua}

 \vskip 6mm

 \noindent\Large\bf\thesection. #1}

 \vskip 3mm}
 \def\subsectionn#1{\refstepcounter{subsection}{\color{myaqua}

 \vskip 5mm

 \noindent\large\bf\thesubsection. #1}

 \vskip 2mm}

 \def\boxx#1#2#3#4#5{
 {\linethickness{#4pt}\put(#1,#5){\color{myaqua}{\line(1,0){#3}}}}
 \multiput(#1,#2)(0,#4){2}{\line(1,0){#3}}
 \multiput(#1,#2)(#3,0){2}{\line(0,1){#4}}
  }

\begin{document}

 \fancyhead[L]{\hspace*{-13mm}\small
 \bt{l}{\bf Journal of High Energy Physics, Gravitation and Cosmology, 2017, \textbf{3}, 368-387}\\
 Published Online in SciRes. \href{http://www.scirp.org/journal/jhepgc}{\color{blue}{\underline{\smash{http://www.scirp.org/journal/jhepgc}}}} \\
 \href{http://dx.doi.org/10.4236/jhepgc.2017.32031}{\color{blue}{\underline{\smash{http://dx.doi.org/10.4236/jhepgc.2017.32031}}}} \\
 \et}

 $\mbox{ }$

 \vskip 12mm

{ 

{\noindent{\huge\bf\color{myaqua}
On the Cohomological Derivation of Yang-Mills Theory in the Antifield Formalism}}
%
\\[6mm]
{\large\bf Ashkbiz Danehkar$^{1,2}$}}
\\[2mm]
{{\small 
$^1$Faculty of Physics, University of Craiova, Craiova, Romania\\
$^2$Present Address: Center for Astrophysics, Cambridge, MA, USA\\
Email: ashkbiz.danehkar@cfa.harvard.edu
 \\[4mm]
Received January 17, 2017, Accepted April 27, 2017, Published April 30, 2017
 \\[4mm]
Copyright \copyright \ 2017 by author and Scientific Research Publishing Inc. \\
This work is licensed under the Creative Commons Attribution International License (CC BY 4.0). \\
\href{http://creativecommons.org/licenses/by/4.0/}{\color{blue}{\underline{\smash{http://creativecommons.org/licenses/by/4.0/}}}}}

\lin{5}{7}

 { 
 {\noindent{\large\bf\color{myaqua} Abstract}{\bf \\[3mm]
 \textup{We present a brief review of the cohomological solutions of self-coupling interactions of the fields in the free Yang-Mills theory. All consistent interactions among the fields have been obtained using the antifield formalism through several order BRST deformations of the master equation. It is found that the coupling deformations halt exclusively at the second order, whereas higher order deformations are obstructed due to non-local interactions. The results demonstrate the BRST cohomological derivation of the interacting Yang-Mills theory.
 }}}
 \\[4mm]
 {\noindent{\large\bf\color{myaqua} Keywords}{\bf \\[3mm]
 Yang-Mills Theory; BRST Symmetry; BRST Cohomology; Antifield Formalism
}

 \fancyfoot[L]{~\newline\footnotesize{\noindent{\color{myaqua}{\bf How to cite this
 paper:}} A. Danehkar (2017)
 On the Cohomological Derivation of Yang-Mills Theory in the Antifield Formalism.
 \textit{Journal of High Energy Physics, Gravitation and Cosmology}, \textbf{3}, 368-387}. \\{\color{blue}{\underline{\smash{http://dx.doi.org/10.4236/jhepgc.2017.32031}}}}}

\lin{3}{1}

\section{Introduction}

Dirac's pioneering approach \cite{Dirac1950,Dirac1958,Dirac1964}
has been used for constrained systems in quantum field theory \cite{Anderson1951,Bergmann1955,Weinberg1995}. This approach allowed us to construct the action in either Lagrangian or Hamiltonian forms \cite{Gotay1979,Gotay1980}, while both of them are equivalent \cite{Batlle1986}. In this way, the Hamiltonian quantization is derived using canonical variables (coordinate and momentum) involving constrained dynamics \cite{Govaerts1990,Govaerts1991,Hanson1976,Landau1976,Sudarshan1974,Sundermeyer1982}. Physical
variables of a constrained system possess gauge invariance and locally independent symmetry. 
The gauge symmetry introduces some arbitrary time independent functions to the Hamilton's equations of motion. 
We notice that all canonical variables are not independent. Therefore, some conditions for canonical variables are required to be imposed, i.e., the first- and second-class constraints. Furthermore, the framework should be generalized to include both commutative (bosonic) and anticommutative (fermionic) variables in constrained systems.

\renewcommand{\headrulewidth}{0.5pt}
\renewcommand{\footrulewidth}{0pt}
 \pagestyle{fancy}
 \fancyfoot{}
 \fancyhead{} 
 \fancyhf{}
 \fancyhead[RO]{\leavevmode \put(-90,0){\color{myaqua}\small A. Danehkar} \boxx{15}{-10}{10}{50}{15} }
 \fancyhead[LE]{\leavevmode \put(0,0){\color{myaqua}\small A. Danehkar} \boxx{-45}{-10}{10}{50}{15} }
 \fancyfoot[C]{\leavevmode
 \put(-2.5,-3){\color{myaqua}\thepage}}
 \renewcommand{\headrule}{\hbox to\headwidth{\color{myaqua}\leaders\hrule height \headrulewidth\hfill}}
 
To generalize constrained systems for canonical conditions and (anti-)commutative variables, Becchi, Rouet, Stora \cite{Becchi1974,Becchi1975,Becchi1976}, and Tyutin \cite{Tyutin1975} developed the BRST formalism to extend the gauge symmetry in terms of the BRST differential and co-/homological classes. The aim was to replace the original gauge symmetry with the BRST symmetry. Noting that the gauge symmetry can be constructed from a nilpotent derivation, so the gauge action is invariant under a nilpotent symmetry, called the BRST symmetry. By replacing the original gauge symmetry with the BRST symmetry, antifield, ghosts, and
antighosts are introduced for each gauge variable \cite{Brandt1989a,Brandt1989}. It yields a generalized framework for solutions of the equations of motion \cite{Fisch1989,Henneaux1990}. Moreover, BRST cohomology extended by the antifield formalism \cite{Batalin1981,Batalin1983,Batalin1984,Fisch1990,Henneaux1990,Henneaux1991,Gomis1993,Gomis1995} allowed us to construct all consistent interactions among the fields using coupling deformations of the master equation \cite{Barnich1993,Henneaux1998}. The BRST-antifield formalism appears as efficient mathematical tool to analyze the consistent interactions, and has been applied to many gauge models, e.g., Yang-Mills model \cite{Barnich1995a}, 
topological Yang-Mills model \cite{Bizdadea2000}, 5-D topological BF model \cite{Cioroianu2005}, and 5-D dual linearized gravity  coupled to topological BF model \cite{Bizdadea2009}.

In this paper, we briefly review the construction of all consistent interactions of
the free Yang--Mills theory determined from 
all coupling deformations of the master
equation. We see that the resulting
action presents deformed structures of the gauge transformation and yields a
commutator for it. In Section~\ref%
{sec2}, the BRST differential and the antifield formalism are introduced. 
Section~\ref{sec3} introduces the consistent interactions among the fields.  
We consider the BRST coupling deformations of the master equations in the antifield
formalism in Section~\ref{sec4}. In Section~\ref{sec5}, we demonstrate its application to the
massless Yang--Mills theory by calculating all several order deformation of the master equation. 
Section~\ref{sec6} presents a conclusion.

\section{BRST Differential}

\label{sec2}

The gauge invariant in a phase space implies that the smooth phase space $%
C^{\infty }(P)$ is substituted by the smooth manifold of the constraint
surface $C^{\infty }(\Sigma )$ while the elements of $C^{\infty }(\Sigma )$
vanish due to the longitudinal exterior derivative on
manifold $\Sigma $. The manifold $\Sigma $, which is embedded in a phase space and a
set of vectors tangent to $\Sigma $, and is closed on it, presents the
definition of the gauge orbits. It manifests the presentation of a nilpotent
derivation $s$, the so-called BRST differential, that includes an
algebra involving $C^{\infty }(P)$, where the cohomology of $s$ indicates
that the gauge transformations of the constraint
surface $C^{\infty }(\Sigma )$ are constant along the gauge orbits (denoted
by $\mathcal{G}$).

The reduced space, by taking $\Sigma $ over gauge orbits, denote by algebra $C^{\infty }(\Sigma /\mathcal{G})$,  
includes all variables of the gauge invariant. However, it is not possible to construct $C^{\infty
}(\Sigma /\mathcal{G})$ from physical observables, as one cannot solve equations defining $\Sigma $ and trace the gauge orbits $\mathcal{G}$.
Hence, the BRST symmetry should be used to reformulate the physical
observables in a convenient approach. To construct the BRST differential $s$, two auxiliary derivations $\delta $ and $\gamma$  are introduced. 
The differential of the first derivation $\delta $ is called the \textit{Koszul-Tate
differential} that yields a resolution of the smooth manifold of the constraint
surface $C^{\infty }(\Sigma )$. The second differential is called the \textit{longitudinal
differential} $\gamma $ along the gauge orbits in such its zeroth cohomology
group provides the functions on the surface $\Sigma $ being constant along
the gauge orbits $\mathcal{G}$. Hence, the BRST differential $s$ is decomposed into \cite%
{Fisch1989,Fisch1990,Henneaux1990} 
\begin{equation}
s=\delta +\gamma ,  \label{eq:e_3_101}
\end{equation}%
whose cohomology is equal to the cohomology of the longitudinal differential 
$\gamma $, while the Koszul-Tate differential $\delta $ restricts it to the constrains surface $C^{\infty }(\Sigma )$. Note that the
BRST symmetry acts as a general odd derivation on the original fields and
some auxiliary fields (antifields and ghosts), which are equipped for any $X$\ and $Y$\ with Grassmann
parity $\varepsilon_{X}$ and $\varepsilon_{Y}$: 
\begin{equation}
\begin{array}{cc}
s(XY)=X(sY)+(-1)^{\varepsilon _{Y}}(sX)Y, & (\mathrm{Leibniz~rule})%
\end{array}
\label{eq:a_4_72}
\end{equation}%
\begin{equation}
\begin{array}{cc}
s^{2}=0. & (\mathrm{nilpotency})%
\end{array}
\label{eq:a_4_73}
\end{equation}%
where ${\varepsilon_{X}}=0$ or $1$ for bosonic (commutative) or fermionic (anticommutative) variable $X$, respectively. 

Any nilpotent derivation has a degree in a $N$-grading space denoted by 
\begin{equation}
\deg (s)=\pm 1.  \label{eq:e_2_99}
\end{equation}%
The positive degree of the differential $s$ increases the grading while the
negative degree decreases it, i.\thinspace e. $s(X_{n})\subset X_{n\pm 1}$
depending on the degree of the differential operator. The grading of $s$ is
the so-called \textit{ghost number} (${\mathfrak{gh}}$), equal to one, 
consists of the pureghost number (${\mathfrak{pgh}}$) and the antighost number (${\mathfrak{agh}}$):
\begin{equation}
\mathfrak{gh}(X)=\mathfrak{pgh}(X)-\mathfrak{agh}(X),
\end{equation}%
with the following property%
\begin{equation}
\mathfrak{gh}(XY)=\mathfrak{gh}(X)+\mathfrak{gh}(Y),
\end{equation}%
where the operators $\mathfrak{pgh}$ and $\mathfrak{agh}$ stand for the
pureghost and antighost numbers, respectively. For the
Koszul-Tate differential $\delta $ and the longitudinal differential $\gamma 
$, we get:
\begin{equation}
\begin{array}{cccc}
{\mathfrak{pgh}(\delta )=0,} & {\mathfrak{agh}(\delta )=-1,} & {\mathfrak{pgh%
}(\gamma )=1,} & {\mathfrak{agh}(\gamma )=0,}%
\end{array}%
\end{equation}%
such $\mathfrak{gh}{(s)}=\mathfrak{gh}{(\delta )}=\mathfrak{gh}{(\gamma )}=1$. 
The differentials $\delta $ and $\gamma $ increase the
ghost number by one unit. The differential $\delta $ reduces the antighost
number, but maintains the pureghost number, whereas the differential $%
\gamma $ increases the pureghost number, but maintains the antighost
number.

The cohomology algebra of the differential $s$ is $H(s)={\mathrm{Ker}~s}/{\mathrm{Im}~s}$, where the elements of the kernel subspace, $\mathrm{Ker}~s$, are closed and vanish via 
the differential $s$:
\begin{equation}
\begin{array}{cc}
{sa=0,} & {a\in \mathrm{Ker}~s,}%
\end{array}
\label{eq:e_2_100}
\end{equation}%
while the elements of its image subspace, $\mathrm{Im}~s$,
are exact:
\begin{equation}
\begin{array}{cc}
{sb=a,} & {a\in \mathrm{Im}~s.}%
\end{array}
\label{eq:e_2_101}
\end{equation}%
The cohomology algebra of $s$, denoted by $H^{k}(s)$ ($k$ is a cohomology degree), exists if its degree is positive, whereas its homology algebra, denoted by $H_{k}(s)$, has a negative degree. The co-/homology
with the grading algebra then reads as follows
\begin{equation}
\begin{array}{l}
\deg (s)=+1\rightarrow H^{k}(s)=\mathop\oplus \limits_{n \in \mathds{N}}H^{n}(s), \\ 
\deg (s)=-1\rightarrow H_{k}(s)=\mathop\oplus \limits_{n \in \mathds{N}}H_{n}(s).%
\end{array}
\label{eq:e_2_103}
\end{equation}%
If the co-/homology $H^{k}(s)$ is zero, the differential $s$ is called to
be acyclic in a degree of $k$.

The zeroth cohomology group of the BRST differential $%
H^{0}(s) $ leads to equation (\ref{eq:a_4_73}), the essential aspect of the
BRST symmetry, that implies the vanishing squares of its
derivations $\delta $ and $\gamma $: 
\begin{equation}
\begin{array}{cc}
{\delta ^{2}=0,} & {\gamma ^{2}=0.}%
\end{array}
\label{eq:e_3_103}
\end{equation}%
and also their anticommutation: 
\begin{equation}
\gamma \delta +\delta \gamma =0.
\end{equation}%
It means that the Koszul-Tate differential $\delta $ commutes with the
longitudinal differential $\gamma $.

The generator of the Koszul-Tate complex may be chosen in an equal number of
freedom as the generator of the longitudinal exterior complex. It follows
that they are canonically conjugate in the extended space of 
original and new generators of $%
\delta $ and $\gamma $. This implies that the BRST transformation
maintains a canonical transformation in the BRST complex space ${%
\mathds{C}[x^k ]}$ through a bracket structure: 
\begin{equation}
\begin{array}{cc}
{sX=[X,\Omega ],} & {\forall X\in \mathds{C}[x^k ],}%
\end{array}
\label{eq:e_3_105}
\end{equation}%
which is called the \textit{Poisson bracket} and defined as follows: 
\begin{equation}
\lbrack X,Y]\equiv \frac{{\partial X}}{{\partial q^{k}}}\frac{{\partial Y}}{{%
\partial p_{k}}}-\frac{{\partial X}}{{\partial p_{k}}}\frac{{\partial Y}}{{%
\partial q^{k}}}  \label{eq:e_2_20}
\end{equation}%
where $q_{k}$ and ${p_{k}}$ are positions and canonical momenta of a Hamiltonian system, respectively. 

Equation (\ref{eq:e_3_105}) represents the BRST symmetry in the
Hamiltonian formalism. The choice of $s$
as canonical transformation manifests the BRST symmetry where the
canonical variables remain unchanged under transformation. The fermionic
charge $\Omega $ is called the BRST generator for the Hamiltonian formalism.
Applying the Jacobi identity to the Poisson bracket and the nilpotency
definition of the BRST differential yields:
\begin{equation}
\lbrack \Omega ,\Omega ]=0,  \label{eq:e_3_106}
\end{equation}%
which is the \textit{master equation} of the BRST generator in the Hamiltonian
formalism.

\section{Consistent Interactions}

\label{sec3}

To understand the consistent interactions among fields with a gauge freedom, we begin our study with a Lagrangian action: 
\begin{equation}
S_{0}^{\mathrm{L}}[{\phi ^{\alpha _{0}}]=}\int d^{D}x\mathcal{L}_{0}\left( {%
\phi ^{\alpha _{0}},\partial }_{\mu }{\phi }^{\alpha _{0}},{\partial }_{\mu }%
{\partial }_{\nu }{\phi }^{\alpha _{0}},{\ldots ,\partial }_{\mu _{1}}{%
\partial }_{\mu _{2}}\cdots {\partial }_{\mu _{k}}{\phi }^{\alpha
_{0}}\right) ,  \label{eq:e_5_1}
\end{equation}%
where the action $S_{0}^{\mathrm{L}}[{\phi ^{\alpha _{0}}]}$ is local
functional of the fields ${\phi ^{\alpha _{0}}}$ and their Lorentz covariant
derivatives. 

The equations of motion then reads 
${\delta }S_{0}^{\mathrm{L}}/{\delta \phi }^{\alpha _{0}}(x) =0,$
where ${\delta }S_{0}^{\mathrm{L}}/{\delta \phi }^{\alpha _{0}}$\ is
functional derivatives. The action $S_{0}^{\mathrm{L}}[{\phi ^{\alpha _{0}}]}$ possesses generic 
free gauge symmetries 
\begin{equation}
\delta _{\varepsilon }\phi ^{\alpha _{0}}=Z_{\alpha _{1}}^{\alpha
_{0}}\varepsilon ^{\alpha _{1}},  \label{eq:e_5_2}
\end{equation}%
The equations of motion is then determined from the action principle: 
$\delta _{\varepsilon }S_{0}^{\mathrm{L}}[\phi ^{\alpha _{0}}]=0$.

Let consider the deformations of the action in such a way 
\begin{equation}
S_{0}^{\mathrm{L}}[\phi ^{\alpha _{0}}]\longrightarrow S^{\mathrm{L}}[\phi
^{\alpha _{0}}]={S_{0}^{\mathrm{L}}}[\phi ^{\alpha _{0}}]+\lambda {S_{1}^{%
\mathrm{L}}}[\phi ^{\alpha _{0}}]+\lambda ^{2}{S_{2}^{\mathrm{L}}}[\phi
^{\alpha _{0}}]+\ldots , \label{eq:e_5_8}
\end{equation}%
that implies the deformation of gauge symmetries as 
\begin{equation}
Z_{\alpha 1}^{\alpha _{0}}\longrightarrow \bar{Z}_{\alpha 1}^{\alpha
_{0}}=Z_{\alpha 1}^{\alpha _{0}}+\lambda \overset{\smash{(1)}}{Z}_{\alpha
_{1}}^{\alpha _{0}}+\lambda ^{2}\overset{\smash{(2)}}{Z}_{\alpha 1}^{\alpha
_{0}}+\ldots .  \label{eq:e_5_9}
\end{equation}%
This provides the deformed gauge transformations: 
\begin{equation}
\frac{{\delta S}^{\mathrm{L}}}{{\delta \phi ^{\alpha _{0}}}}\bar{Z}_{\alpha
_{1}}^{\alpha _{0}}=0.  \label{eq:e_5_10}
\end{equation}%
Equations (\ref{eq:e_5_8}) and (\ref{eq:e_5_9}) lead to the following expression: 
\begin{equation}
\left( \frac{\delta {S_{0}^{\mathrm{L}}}}{\delta \phi ^{\alpha _{0}}}%
+\lambda \frac{\delta {S_{1}^{\mathrm{L}}}}{\delta \phi ^{\alpha _{0}}}%
+\lambda ^{2}\frac{\delta {S_{2}^{\mathrm{L}}}}{\delta \phi ^{\alpha _{0}}}%
+\ldots \right) \left( Z_{\alpha _{1}}^{\alpha _{0}}+\lambda \overset{%
\smash{(1)}}{Z}_{\alpha _{1}}^{\alpha _{0}}+\lambda ^{2}\overset{\smash{(2)}}%
{Z}_{\alpha _{1}}^{\alpha _{0}}+\ldots \right) =0.  \label{eq:e_5_11}
\end{equation}%
Hence, the deformations by their orders are as follows:
\begin{equation}
\left\{ 
\begin{array}{cc}
\lambda ^{0}: & \displaystyle \frac{{\delta S_{0}^{\mathrm{L}}}}{{\delta
\phi ^{\alpha _{0}}}}Z_{\alpha _{1}}^{\alpha _{0}}=0, \\ 
\lambda ^{1}: & \displaystyle \frac{{\delta S_{0}^{\mathrm{L}}}}{{\delta
\phi ^{\alpha _{0}}}}\overset{\smash{(1)}}{Z}_{\alpha _{1}}^{\alpha _{0}}+%
\displaystyle \frac{{\delta S_{1}^{\mathrm{L}}}}{{\delta \phi ^{\alpha _{0}}}%
}Z_{\alpha _{1}}^{\alpha _{0}}=0, \\ 
\lambda ^{2}: & \displaystyle \frac{{\delta S_{0}^{\mathrm{L}}}}{{\delta
\phi ^{\alpha _{0}}}}\overset{\smash{(2)}}{Z}_{\alpha _{1}}^{\alpha _{0}}+%
\displaystyle \frac{{\delta S_{1}^{\mathrm{L}}}}{{\delta \phi ^{\alpha _{0}}}%
}\overset{\smash{(1)}}{Z}_{\alpha _{1}}^{\alpha _{0}}+\displaystyle \frac{{%
\delta S_{2}^{\mathrm{L}}}}{{\delta \phi ^{\alpha _{0}}}}Z_{\alpha
_{1}}^{\alpha _{0}}=0, \\ 
\vdots & \vdots%
\end{array}%
\right.  \label{eq:e_5_12}
\end{equation}%
which define the deformed gauge transformations that close
on-shell for the interacting action, the so-called \textit{consistent
interactions},\ while the original gauge transformations are reducible \cite{Henneaux1991}.

Assume that the gauge fields of consistent interactions are trivially defined to be the following sum:  
\begin{equation}
\phi ^{\alpha _{0}}\longrightarrow \bar{\phi}^{\alpha _{0}}=\phi ^{\alpha
_{0}}+\lambda F^{\alpha _{0}}[\phi ^{\beta _{0}}]+\lambda ^{2}F^{\alpha
_{0}}[\phi ^{\beta _{0}}]+\ldots,  \label{eq:e_5_13}
\end{equation}%
we then obtain 
\begin{eqnarray}
{S_{0}^{\mathrm{L}}}[\phi ^{\alpha _{0}}]\longrightarrow {S}^{\mathrm{L}%
}[\phi ^{\alpha _{0}}] &=&{S_{0}^{\mathrm{L}}}[\phi ^{\alpha _{0}}+\lambda
F^{\alpha _{0}}+\ldots ]  \notag \\
&=&{S_{0}^{\mathrm{L}}}[\phi ^{\alpha _{0}}]+\lambda {\frac{\delta {S_{0}^{%
\mathrm{L}}}}{\delta \phi ^{\alpha _{0}}}}F_{1}^{\alpha _{0}}  \notag \\
&&+\lambda ^{2}\left( {\frac{\delta ^{2}{S_{0}^{\mathrm{L}}}}{\delta \phi
^{\alpha _{0}}\delta \phi ^{\beta _{0}}}}F_{1}^{\alpha _{0}}F_{1}^{\beta
_{0}}+{\frac{\delta ^{2}{S_{0}^{\mathrm{L}}}}{\delta (\phi ^{\alpha
_{0}})^{2}}}F_{2}^{\alpha _{0}}\right) \ +\ldots,  \label{eq:e_5_14}
\end{eqnarray}%
which does not manifest an exact interacting theory. A theory
is strict if the consistent deformations are merely proportional to its free theory action ${S_{0}^{\mathrm{L}}}[\phi ^{\alpha _{0}}]$ up to the
redefinition of the gauge fields. Thus, the interaction is formulated
as follows:
\begin{equation*}
{S_{0}^{\mathrm{L}}}[\phi ^{\alpha _{0}}]\longrightarrow {S}^{\mathrm{L}%
}[\phi ^{\alpha _{0}}]=\left( 1+\mathcal{Q}_{1}\lambda +(\mathcal{Q}%
_{2}\lambda )^{2}+\ldots \right) {S_{0}^{\mathrm{L}}}[\phi ^{\alpha _{0}}] 
\end{equation*}%
where charges $\mathcal{Q}_{k}$ in the $k$ order of the coupling constants $%
\lambda ^{k}$ are given by 
\begin{equation}
\begin{array}{c}
\mathcal{Q}_{1}\equiv {\displaystyle \frac{\delta }{\delta \phi ^{\alpha
_{0}}}}F_{1}^{\alpha _{0}}, \\ 
\mathcal{Q}_{2}^{2}\equiv {\displaystyle \frac{\delta ^{2}}{\delta \phi
^{\alpha _{0}}\delta \phi ^{\beta _{0}}}}F_{1}^{\alpha _{0}}F_{1}^{\beta
_{0}}+{\displaystyle \frac{\delta ^{2}}{\delta (\phi ^{\alpha _{0}})^{2}}}%
F_{2}^{\alpha _{0}}, \\ 
\vdots%
\end{array}
\label{eq:e_5_16}
\end{equation}%
It represents the unperturbed action by charges of the coupling constants.

\section{BRST Deformations of the Master Equation}

\label{sec4}

Let us consider the gauge transformation defined by the equation (\ref{eq:e_5_2}%
). The classical fields $\phi ^{\alpha _{0}}$ possesses the ghost number
zero. It implies an ghost $\eta ^{\alpha _{1}}$ associated to ghost number
one, as well as the one-level ghost of ghost $\eta ^{\alpha _{2}}$ have
number two, etc, i.\thinspace e. 
\begin{equation}
\eta {^{A}=}\left\{ \eta ^{\alpha _{1}}{,\ldots ,}\eta ^{\alpha
_{k}}\right\} ,  \label{eq:e_5_17_1}
\end{equation}%
which have the following ghost numbers, $\mathfrak{gh}$, and Grassmann parities, $\varepsilon$: 
\begin{equation}
\begin{array}{cc}
{\mathfrak{gh}(\eta ^{\alpha _{k}})=k,} & {\varepsilon (\eta ^{\alpha
_{k}})=k\mathrm{~(mod~2)}.}%
\end{array}
\label{eq:e_5_17_2}
\end{equation}%
It also implies antifields $\phi _{\alpha _{0}}^{\ast }$ and antighosts $%
\eta _{A}^{\ast }$\ of opposite Grassmann parity with the following ghost
numbers, $\mathfrak{gh}$, and Grassmann parities, $\varepsilon$, respectively:
\begin{equation}
\begin{array}{cc}
{\mathfrak{gh}(\phi _{\alpha _{0}}^{\ast })=-\mathfrak{gh}(\phi ^{\alpha
_{0}})-1,} & {\varepsilon (\phi _{\alpha _{0}}^{\ast })=\varepsilon (\phi
^{\alpha _{0}})+1\mathrm{~(mod~2),}}%
\end{array}
\label{eq:e_5_17_3}
\end{equation}%
\begin{equation}
\begin{array}{cc}
{\mathfrak{gh}(\eta _{\alpha _{k}}^{\ast })=-(k+1),} & {\varepsilon (\eta
_{\alpha _{k}}^{\ast })=k+1\mathrm{~(mod~2)}.}%
\end{array}
\label{eq:e_5_17_4}
\end{equation}%
The presentation of the gauge variables is therefore provided by 
\begin{equation}
\begin{array}{cc}
{\Phi ^{A}=\left\{ {\phi ^{\alpha _{0}},\eta ^{A}}\right\} ,} & {\Phi
_{A}^{\ast }=\left\{ {\phi }_{\alpha _{0}}^{\ast }{,}\eta _{A}^{\ast
}\right\} ,}%
\end{array}
\label{eq:e_5_17_5}
\end{equation}%
where a set of fields ${\Phi }^{A}$ includes the original fields, the
ghost, and the ghosts of ghosts, and ${\Phi _{A}^{\ast }}$ includes the
their corresponding antifields.

The BRST symmetry is a canonical transformation, and defined by an
antibracket structure: 
\begin{equation}
\bar{s}X\equiv (X,S),  \label{eq:a_4_76}
\end{equation}%
where $S$ is the canonical generators, and the antibracket (see appendix~\ref%
{app_1}) is defined in the space of fields ${\Phi ^{A}}$\ and antifields ${%
\Phi _{A}^{\ast }}$ as follows \cite{Batalin1981}: 
\begin{equation}
(X,Y)\equiv \frac{{\partial _{r}X}}{{\partial \Phi ^{A}}}\frac{{\partial
_{l}Y}}{{\partial \Phi _{A}^{\ast }}}-\frac{{\partial _{r}X}}{{\partial \Phi
_{A}^{\ast }}}\frac{{\partial _{l}Y}}{{\partial \Phi ^{A}}}.
\label{eq:e_4_6}
\end{equation}%
The Grassmann parity and ghost number of the antibracket are, respectively: 
\begin{equation}
\varepsilon (X,Y)=\varepsilon _{X}+\varepsilon _{Y}+1\mathrm{~(mod~2)},
\end{equation}%
\begin{equation}
\mathfrak{gh}(X,Y)=\mathfrak{gh}(X)+\mathfrak{gh}(Y)+1.  \label{eq:e_4_13}
\end{equation}%
The antifields are now considered as mathematical tool to
construct the BRST formalism. The solution can be interpreted as source coefficient
for BRST transformation, i.e., an effective action in the theory.

The fields and antifields establish the solution $S[{\Phi }^{A},{\Phi }%
_{A}^{\ast }]$ of the classical master equation for consistent interactions 
\cite{Barnich1993},%
\begin{equation}
{S}={S}_{0}+\lambda {S}_{1}+\lambda ^{2}{S}_{2}+\ldots .  \label{eq:e_5_17_6}
\end{equation}%
Section \ref{sec2} presented the master equation (\ref{eq:e_3_106}) of the
BRST generator in the Hamiltonian formalism. The gauge structure is now
constructed through the solution $S$ of the \textit{master equation} in
the antifield formalism by \cite{Batalin1981,Batalin1983,Barnich1993} 
\begin{equation}
(S,S)=0.  \label{eq:e_5_17_7}
\end{equation}%
This shows the consistency of the gauge transformations. The master
equation (\ref{eq:e_5_17_7}) includes the closure of the gauge transformations, the
higher-order gauge identities, and the Noether identities. The master
equation maintains the consistent specifications on $S_{0}$ and $Z_{\alpha
_{1}}^{\alpha _{0}}$.

Substituting the definition (\ref{eq:e_5_17_6}) into the master
equation (\ref{eq:e_5_17_7}) yields
\begin{equation}
(S_{0}+\lambda S_{1}+\lambda ^{2}S_{2}+\ldots ,S_{0}+\lambda S_{1}+\lambda
^{2}S_{2}+\ldots )=0.  \label{eq:e_5_17_8}
\end{equation}%
We then derive 
\begin{equation}
\left\{ 
\begin{array}{cc}
\lambda ^{0}: & (S_{0},S_{0})=0, \\ 
\lambda ^{1}: & (S_{0},\lambda S_{1})+(\lambda S_{1},S_{0})=0, \\ 
\vdots & \vdots%
\end{array}%
\right.  \label{eq:e_5_17_9}
\end{equation}%
which are simplified as follows \cite{Barnich1993,Bizdadea2009,Barnich1994a,Bizdadea2003} 
\begin{eqnarray}
(S_{0},S_{0}) &=&0,  \label{eq:e_5_17_10} \\
2(S_{0},S_{1}) &=&0,  \label{eq:e_5_17_11} \\
2(S_{0},S_{2})+(S_{1},S_{1}) &=&0,  \label{eq:e_5_17_12} \\
(S_{0},S_{3})+(S_{1},S_{2}) &=&0,  \label{eq:e_5_17_12_1} \\
2\left( S_{0},S_{4}\right) +2\left( S_{1},S_{3}\right) +\left(
S_{2},S_{2}\right) &=&0,  \label{eq:e_5_17_12_2} \\
\left( S_{0},S_{5}\right) +\left( S_{1},S_{4}\right) +\left(
S_{2},S_{3}\right) &=&0,  \label{eq:e_5_17_12_3} \\
&&\vdots \ \ \   \notag
\end{eqnarray}%
the so-called \textit{deformations of the master equation} \cite%
{Barnich1993,Henneaux1998}. 

The equation  (\ref{eq:e_5_17_11}) implies that $S_{1}$ is a cocycle for the free
differential defined by $s\equiv (\cdot ,S_{0})$, i.e., $S_{1}$ is
a coboundary, $S_{1}=(B_{1},S_{0})$. The equation (\ref{eq:e_5_17_10}) hence corresponds to $s^2=0$. 
The equation (\ref{eq:e_5_17_12})
indicates that $(S_{1},S_{1})$ is trivial in $H^{1}(s)$, and $H^{0}(s)$ is mapped trivially into $H^{1}(s)$ by
the antibracket. Furthermore, the higher orders $H^{0}(s)$ 
mapped into $H^{1}(s)$ are trivial, and provide the existence of the
 terms $S_{3},S_{4},$ etc, up to an element of $%
H^{0}(s) $. So, the $k$ orders $\lambda ^{k}$ freely link the
interaction of an arbitrary element of $H^{0}(s)$.

The free gauge invariant action ${S_{0}^{\mathrm{L}}}$ and the gauge
transformations can be retrieved from 
\begin{equation}
S_{0}={S_{0}^{\mathrm{L}}}+\phi _{\alpha _{0}}^{\ast }Z_{\alpha
_{1}}^{\alpha _{0}}\eta ^{\alpha _{1}}+\ldots , \label{eq:e_5_18_1}
\end{equation}%
by setting 
\begin{equation}
{S_{0}^{\mathrm{L}}}=S_{0}[{\Phi }^{A},{\Phi }_{A}^{\ast }=0].
\label{eq:e_5_18_2}
\end{equation}%
It provides the solution $S_{0}$ of the classical master equation for 
\textit{field gauge symmetries},%
\begin{equation}
(S_{0},S_{0})=0.  \label{eq:e_5_18_3}
\end{equation}%
The BRST differential $s$ is now defined by $%
S_{0}$ through the antibracket,%
\begin{equation}
sX\equiv (X,S_{0}).  \label{eq:e_5_18_5}
\end{equation}%
Using the definitions (\ref{eq:e_5_18_5}), the deformations of the
master equation are rewritten as follows: 
\begin{equation}
\begin{array}{cc}
\lambda ^{1}: & 2sS_{1}=0, \\ 
\lambda ^{2}: & (S_{1},S_{1})+2sS_{2}=0, \\ 
\lambda ^{3}: & (S_{1},S_{2})+sS_{3}=0, \\ 
\lambda ^{4}: & 2\left( S_{1},S_{3}\right) +\left( S_{2},S_{2}\right)
+2sS_{4}=0, \\ 
\lambda ^{5}: & \left( S_{1},S_{4}\right) +\left( S_{2},S_{3}\right)
+sS_{5}=0, \\ 
\vdots & \vdots%
\end{array}
\label{eq:e_5_25}
\end{equation}%
which are the deformations of the master equation in terms
of the BRST differential $s$.

\section{BRST Cohomology of the Free Yang-Mills Theory}

\label{sec5}

Let us consider a set of $N$ potentials $A_{\mu }^{a}$
described by the abelian action in terms of the free (massless) Lagrangian action 
\begin{equation}
{S_{0}^{\mathrm{L}}}[A_{\mu }^{a}]=\int d^{D}x\left( -{{\textstyle \frac{1}{4%
}}}F_{\mu \nu }^{a}F_{a}^{\mu \nu }\right) ,\;\;a=1,\ldots ,N,\;\;N\in %
\mathds{N},  \label{eq:e_5_31}
\end{equation}%
where $A_{\mu }^{a}$ is the abelian field potential, $D$\ is the spacetime
dimension, strictly $D>2$, since the theory has no local degree of freedom
in two dimensions, and the abelian field strengths $F_{\mu \nu }^{a}$ is
defined by 
\begin{equation}
F_{\mu \nu }^{a}\equiv \partial _{\mu }A_{\nu }^{a}-\partial _{\nu }A_{\mu
}^{a}=\frac{\partial A_{\nu }^{a}}{\partial x^{\mu }}-\frac{\partial A_{\mu
}^{a}}{\partial x^{\nu }},  \label{eq:e_5_32}
\end{equation}%
in such a way 
\begin{equation}
F_{a}^{\mu \nu }=\sigma ^{\mu \alpha }\sigma ^{\nu \beta }k_{ab}F_{\alpha
\beta }^{b},  \label{eq:e_5_33}
\end{equation}%
where $\sigma ^{\mu \alpha }=\mathrm{diag}(-1,1,\ldots ,1)$ is the $\mathrm{%
SO}(1,D-1)$ invariant flat metric in Minkowski space with the particular
hermitian representation of the Clifford algebra $\left\{ \gamma ^{\mu
},\gamma ^{\nu }\right\} =2\sigma ^{\mu \nu }$, and $k_{ab}$\ is a given
symmetric invertible matrix with following properties 
\begin{equation}
\begin{array}{ccc}
k_{(ab)}=k_{ab}=k_{ba}, & k^{ab}k_{bc}=\delta ^{a}{}_{c}, & a,b,c=1,\ldots
,N.%
\end{array}
\label{eq:e_5_34}
\end{equation}

The gauge transformation with the free equation of motion, 
\begin{equation}
\frac{\delta {S_{0}^{\mathrm{L}}}}{\delta A_{\mu }^{a}}=\partial _{\nu
}F_{a}^{\nu \mu }=0,  \label{eq:e_5_35}
\end{equation}%
manifests an irreducible transformation by 
\begin{equation}
\delta _{\varepsilon }A_{\mu }^{a}=\partial _{\mu }\varepsilon ^{a},
\label{eq:e_5_36}
\end{equation}%
while 
\begin{equation}
\delta _{\varepsilon }F_{\mu \nu }^{a}=\partial _{\mu }\partial _{\nu
}\varepsilon ^{a}-\partial _{\nu }\partial _{\mu }\varepsilon ^{a}=0.
\label{eq:e_5_37}
\end{equation}%
The differential operator $\partial _{\mu }$ is determined by the structure $%
Z_{\alpha _{1}}^{\alpha _{0}}$\ of the gauge transformations of an abelian algebra. The
action (\ref{eq:e_5_31}) is close according to an abelian algebra, and
invariant under the gauge transformation (\ref{eq:e_5_36}). The gauge
invariant (\ref{eq:e_5_36}) eliminates unphysical terms, i.\thinspace e. the
longitudinal and temporal degrees of freedom.

The implementation of the BRST transformation in the
minimal sector provides the field $A_{\mu }^{a}$, its ghost $\eta ^{a}$, and
their antifields $A_{a}^{\ast \mu }$ and $\eta _{a}^{\ast }$ with the
respective Grassmann parities, antighost, pureghost, and (total) ghost
numbers, 
\begin{equation}
\begin{array}{c|c|c|c|c}
Z & A_{\mu }^{a} & A_{a}^{\ast \mu } & \eta ^{a} & \eta _{a}^{\ast } \\ 
\hline
\varepsilon (Z) & 0 & 1 & 1 & 0 \\ 
{\mathfrak{agh}}(Z) & 0 & 1 & 0 & 2 \\ 
{\mathfrak{pgh}}(Z) & 0 & 0 & 1 & 0 \\ 
\rule{0em}{3ex}{\mathfrak{gh}}(Z) & 0 & -1 & 1 & -2%
\end{array}
\label{eq:e_5_41}
\end{equation}%
which can schematically be illustrated: 
\begin{equation*}
\xymatrix{ \varepsilon=0 & \mathop{A^a_\mu}\limits_{{\rm{gh =
0}}}\ar[d] \ar[dr] \ar[r]^{\delta_\varepsilon} & \mathop{\partial
_{\mu }\varepsilon^a}\limits_{~} \ar[d]\\ \varepsilon=1 &
\mathop{A_a^{*\mu}}\limits_{{\rm{gh = -1}}} &
\mathop{\eta^a}\limits_{{\rm{gh = 1}}} \ar[d] \\ \varepsilon=0 & &
\mathop{\eta_a^*}\limits_{{\rm{gh = -2}}}} 
\end{equation*}

We calculate the BRST-differential $s$\ that decomposes into the sum of two
differentials, the Koszul-Tate differential $\delta $ and the longitudinal
differential $\gamma $\ along the gauge orbits. Both $\delta $ and $\gamma $
are derivations, and commute with $\partial _{\mu }$, and acting on $A_{\mu
}^{a}$, $A_{a}^{\ast \mu }$, $\rule{0em}{3ex}\eta ^{a}$, and $\eta
_{a}^{\ast }$ via \cite{Barnich2000,Barnich1995a} 
\begin{equation*}
\begin{array}{c|c|c}
Z & \delta Z & \gamma Z \\ \hline
\rule{0em}{3ex}A_{\mu }^{a} & 0 & \partial _{\mu }\eta ^{a} \\ 
A_{a}^{\ast \mu } & -\displaystyle \frac{\delta {S_{0}^{\mathrm{L}}}}{\delta
A_{\mu }^{a}}=-\partial _{\nu }F_{a}^{\nu \mu } & 0 \\ 
\rule{0em}{3ex}\eta ^{a} & 0 & 0 \\ 
\eta _{a}^{\ast } & -\partial _{\mu }A_{a}^{\ast \mu } & 0%
\end{array}
\end{equation*}
The classical master equation (\ref{eq:e_5_18_3}) of the action (%
\ref{eq:e_5_31}) holds the minimal solution (\ref{eq:e_5_18_1}) in such a
way 
\begin{equation}
S_{0}={S_{0}^{\mathrm{L}}}[A_{\mu }^{a}]+\int d^{D}xA_{a}^{\ast \mu
}\partial _{\mu }\eta ^{a}.  \label{eq:e_5_44}
\end{equation}

\subsection{First-order Deformation}
\label{sec5_1}

We now consider the deformed solution of the master equation for the action 
(\ref{eq:e_5_31}) smoothly in the
coupling constant $\lambda $ that brings to the solution (\ref{eq:e_5_44}),
while the coupling constant $\lambda $ vanishes. In Section \ref{sec4}, we noticed that
the first-order deformation ($\lambda ^{1}$) of the master equation
satisfies the solution $sS_{1}=0$, where $S_{1}$ is bosonic (commutative) function with
ghost number zero.

Let us assume 
\begin{equation}
S_{1}=\int d^{D}x\,a,  \label{eq:e_5_45}
\end{equation}%
where $a$ is a local function. Then, the first-order deformation, $sS_{1}=0$, takes the local form%
\begin{equation}
\int d^{D}x\,sa=0\rightarrow sa=(\delta a+\gamma a)=\partial _{\mu }j^{\mu }
\label{eq:e_5_46}
\end{equation}%
\begin{equation}
{\mathfrak{gh}}\left( a\right) =0,\quad \varepsilon \left( a\right) =0,
\label{eq:e_5_47}
\end{equation}%
where $j^{\mu }$\ is a local current that manifests the non-integrated
density of the first-order deformation corresponding to the local
cohomology of $s$ in ghost number zero, $a\in H^{0}\left( s|d\right)$, where
$d$ is the exterior spacetime differential.

To evaluate Equation (\ref{eq:e_5_46}), we assume
\begin{equation}
\begin{array}{ccccc}
a=\sum\limits_{i=0}^{I}a_{i}, & {\mathfrak{agh}}\left( a_{i}\right) =i, & {%
\mathfrak{gh}}\left( a_{i}\right) =0, & \varepsilon \left( a_{i}\right) =0,
& \forall i=0,\ldots ,I,%
\end{array}
\label{eq:e_5_48}
\end{equation}%
\begin{equation}
\begin{array}{cccc}
j^{\mu }=\sum\limits_{i=0}^{I}\overset{(i)}{{j}^{\mu }}, & {\mathfrak{agh}}(%
\overset{(i)}{{j}^{\mu }})=i, & {\mathfrak{gh}}(\overset{(i)}{{j}^{\mu }})=0,
& \varepsilon (\overset{(i)}{{j}^{\mu }})=0,%
\end{array}
\label{eq:e_5_49}
\end{equation}%
where $\overset{(k)}{{j}^{\mu }}$ are some local currents. Substituting
(\ref{eq:e_5_48}) and (\ref{eq:e_5_49}) into (\ref{eq:e_5_46}) yields%
\begin{equation}
\sum\limits_{i=0}^{I}\delta a_{i}+\sum\limits_{i=0}^{I}\gamma
a_{i}=\sum\limits_{i=0}^{I}\overset{(i)}{\partial _{\mu }{j}^{\mu }},
\label{eq:e_5_50}
\end{equation}%
obviously 
\begin{equation}
\begin{array}{cc}
{\mathfrak{agh}}(\delta a_{i})=i-1, & {\mathfrak{agh}}(\gamma a_{i})=i.%
\end{array}
\label{eq:e_5_51}
\end{equation}%
They can be decomposed on the several orders of the antighost number: 
\begin{equation}
\begin{array}{c|cc}
{\mathfrak{agh}}(Z) & Z &  \\ \hline
I & \gamma a_{I}=\partial _{\mu }\overset{(I)}{{j}^{\mu }}, &  \\ 
I-1 & \delta a_{I}+\gamma a_{I-1}=\partial _{\mu }\overset{(I-1)}{{j}^{\mu }}%
, &  \\ 
k & \delta a_{k+1}+\gamma a_{k}=\partial _{\mu }\overset{(k)}{{j}^{\mu }}, & 
\quad k=0,\ldots ,I-2%
\end{array}
\label{eq:e_5_52}
\end{equation}%
The positive antighost number are strictly given as replacement for
the first expression \cite{Cioroianu2005}: 
\begin{equation}
\begin{array}{cc}
\gamma a_{I}=0,\quad I>0 & \rightarrow a_{I}\in H^{I}(\gamma ).%
\end{array}
\label{eq:e_5_53}
\end{equation}%
To proof it, let us consider $e^{I}$ as the elements with pureghost number $I$
of a basis in the polynomial space. The generic solution of (\ref{eq:e_5_53}%
) then takes the form 
\begin{equation}
a_{I}=\alpha _{I}e^{I},  \label{eq:e_5_59}
\end{equation}%
while%
\begin{equation}
\begin{array}{cc}
{\mathfrak{agh}}(\alpha _{I})=I, & {\mathfrak{pgh}}(e^{I})=I.%
\end{array}
\label{eq:e_5_60}
\end{equation}%
The objects $\alpha _{I}$ obviously are nontrivial in $H^{0}\left( \gamma
\right) ,$ the so-called invariant polynomials. In other words, the strict
positive antighost numbers provide trivially the cohomology of the exterior
differential $\gamma $ in the space of invariant polynomials $\alpha _{I}$.
Hence, $\gamma a=\partial _{\mu }{j}^{\mu }$ reduces to $\gamma
a=0 $ (see \cite{Cioroianu2005} for general proof).

Moreover, $a_{I}$ may exclusively be reduced to $\gamma $-exact terms 
\begin{equation}
a_{I}=\gamma b_{I},  \label{eq:e_5_54}
\end{equation}%
corresponding to a trivial definition, which states $a_{I}=0$. This result
is obviously given by the second-order nilpotency of $\gamma $ that
implies the unique solution of (\ref{eq:e_5_53}) up to $\gamma $-exact
contributions, i.\thinspace e. 
\begin{equation}
a_{I}\rightarrow a_{I}+\gamma b_{I},  \label{eq:e_5_55}
\end{equation}%
\begin{equation}
\begin{array}{ccc}
{\mathfrak{agh}}\left( b_{I}\right) =I, & {\mathfrak{pgh}}\left(
b_{I}\right) =I-1, & \varepsilon \left( b_{I}\right) =1.%
\end{array}
\label{eq:e_5_56}
\end{equation}%
Hence, the non-triviality of the first-order deformation $a_{I}$ requires
the cohomology of the exterior longitudinal derivative $\gamma $ in
pureghost number equal to $I$, i.\thinspace e. $a_{I}\in H^{I}(\gamma )$. To
solve (\ref{eq:e_5_52}), it is necessary to provide the cohomology of $\gamma 
$ and $\delta $ , $H\left( \gamma \right) $ and $H\left( \delta |d\right) $: 
\begin{equation}
\begin{array}{cc}
\delta a_{I}=\partial _{\mu }m_{I}^{\mu } & \rightarrow a_{I}\in
H_{I}(\delta |d),%
\end{array}
\label{eq:e_5_58}
\end{equation}%
where%
\begin{equation}
H_{I}(\delta |d)=\{a~|~{\mathfrak{agh}}\left( a\right) =I,~\delta a=\partial
_{\mu }m^{\mu }\}/N.  \label{eq:e_5_57}
\end{equation}

For an irreducible linear situation, where gauge generators are field
independent, we assume that%
\begin{equation}
H_{I}(\delta |d)=0,\quad I>2.  \label{eq:e_5_61}
\end{equation}%
where $H_{I}\left( \delta |d\right) $ manifests the local cohomology of the
Koszul-Tate differential $\delta $, while antighost number is $I$ and
pureghost number vanishes. In this case ($I=2$), we obtain 
\begin{equation}
\left\{ 
\begin{array}{c}
\gamma a_{2}=0, \\ 
\delta a_{2}+\gamma a_{1}=\partial _{\mu }\overset{(1)}{{j}^{\mu }}, \\ 
\delta a_{1}+\gamma a_{0}=\partial _{\mu }\overset{(0)}{{j}^{\mu }}.%
\end{array}%
\right.  \label{eq:e_5_62}
\end{equation}%
The first-order deformation up to antighost number two are:
\begin{equation}
a=a_{0}+a_{1}+a_{2}.  \label{eq:e_5_63}
\end{equation}%
The $a_{2}$ is generated by arbitrarily smooth functions in the form (\ref%
{eq:e_5_59}), with $\alpha _{2}$ from $H_{2}^{\text{inv}}\left( \delta
|d\right) $ and $e^{2}$ denote the elements with pureghost number two of a
basis in the polynomial space, i.e.,
\begin{equation}
\begin{array}{cc}
a_{2}\in H_{2}^{\text{inv}}\left( \delta |d\right) \rightarrow {\mathfrak{agh%
}}(\alpha _{2})=2, & {\mathfrak{pgh}}(e^{2})=2,%
\end{array}
\label{eq:e_5_64}
\end{equation}%
where $H_{I}^{\text{inv}}\left( \delta |d\right) $ is the local cohomology
of the Koszul-Tate differential $\delta $\ with antighost number $I$ in the
invariant polynomial space.

We now consider the Koszul-Tate differential $\delta $ and the exterior
longitudinal differential $\gamma $\ in the action (\ref%
{eq:e_5_44}):%
\begin{equation*}
\begin{array}{ccc}
\delta A_{\mu }^{a}=\delta \eta ^{a}=0, & \delta A_{a}^{\ast \mu }=-\partial
_{\nu }F_{a}^{\nu \mu }, & \delta \eta _{a}^{\ast }=-\partial _{\mu
}A_{a}^{\ast \mu },%
\end{array}%
\end{equation*}%
\begin{equation*}
\begin{array}{cc}
\gamma A_{\mu }^{a}=\partial _{\mu }\eta ^{\alpha }, & \gamma A_{a}^{\ast
\mu }=\gamma \eta ^{a}=\gamma \eta _{a}^{\ast }=0.%
\end{array}%
\end{equation*}%
The local cohomology of the exterior longitudinal derivative $\gamma $ in
pureghost number one, $H^{1}(\gamma ),$ has one ghost $\eta ^{a}$,
while $H^{2}(\gamma )$\ has two ghosts $\eta ^{a}\eta ^{b}$, i.\thinspace e. 
\begin{equation}
\begin{array}{cc}
\{\eta ^{a}\}\in H^{1}(\gamma ), & \{\eta ^{a}\eta ^{b}\}\in H^{2}(\gamma )%
\end{array}
\label{eq:e_5_67}
\end{equation}%
From (\ref{eq:e_5_67}), we then solve 
\begin{equation*}
\gamma a_{2}=0,
\end{equation*}%
by 
\begin{equation}
a_{2}={{\textstyle \frac{1}{2}}}\eta _{a}^{\ast }f_{bc}^{a}\rule{0em}{3ex}%
\eta ^{b}\rule{0em}{3ex}\eta ^{c},  \label{eq:e_5_68}
\end{equation}%
where $f_{bc}^{a}$ contains the structure constants of a non-abelian algebra coupling the Yang-Mills fields, and it is antisymmetric on indices $bc$: 
\begin{equation}
f_{bc}^{a}=f_{[bc]}^{a}\rightarrow f_{bc}^{a}=-f_{cb}^{a}.  \label{eq:e_5_69}
\end{equation}

The expression $\delta a_{2}+\gamma a_{1}=\partial _{\mu }\overset{(1)}%
{{j}^{\mu }}$ is solved by taking the Koszul-Tate differential $\delta $\ from (\ref%
{eq:e_5_68}):%
\begin{eqnarray}
\delta a_{2} &=&{{\textstyle \frac{1}{2}}}\delta (\eta _{a}^{\ast }f_{bc}^{a}%
\rule{0em}{3ex}\eta ^{b}\rule{0em}{3ex}\eta ^{c})  \notag \\
&=&-{{\textstyle \frac{1}{2}}}\partial _{\mu }(A_{a}^{\ast \mu }f_{bc}^{a}%
\rule{0em}{3ex}\eta ^{b}\rule{0em}{3ex}\eta ^{c})+\gamma (A_{a}^{\ast \mu
}f_{bc}^{a}\rule{0em}{3ex}\eta ^{b}A_{\mu }^{c})  \label{eq:e_5_70}
\end{eqnarray}%
We simply notice that 
\begin{equation}
\delta a_{2}-\gamma (A_{a}^{\ast \mu }f_{bc}^{a}\rule{0em}{3ex}\eta
^{b}A_{\mu }^{c})=-{{\textstyle \frac{1}{2}}}\partial _{\mu }(A_{a}^{\ast
\mu }f_{bc}^{a}\rule{0em}{3ex}\eta ^{b}\rule{0em}{3ex}\eta ^{c}).
\label{eq:e_5_71}
\end{equation}%
This indicates 
\begin{equation}
\begin{array}{cc}
a_{1}=-A_{a}^{\ast \mu }f_{bc}^{a}\rule{0em}{3ex}\eta ^{b}A_{\mu }^{c}, & 
\overset{(1)}{{j}^{\mu }}=-\frac{1}{2}A_{a}^{\ast \mu }f_{bc}^{a}\rule%
{0em}{3ex}\eta ^{b}\rule{0em}{3ex}\eta ^{c}.%
\end{array}
\label{eq:e_5_72}
\end{equation}

To obtain $a_{0}$, we solve $\delta a_{1}+\gamma a_{0}=\partial _{\mu
}\overset{(0)}{{j}^{\mu }}$ by taking the Koszul-Tate differential $\delta $%
\ from $a_{1}$: 
\begin{eqnarray}
\delta a_{1} &=&\delta (-A_{a}^{\ast \mu }f_{bc}^{a}\rule{0em}{3ex}\eta
^{b}A_{\mu }^{c})  \notag \\
&=&\partial _{\nu }(-F_{a}^{\nu \mu }f_{bc}^{a}\rule{0em}{3ex}\eta
^{b}A_{\mu }^{c})+\gamma ({{\textstyle \frac{1}{2}}}F_{a}^{\nu \mu
}f_{bc}^{a}A_{\nu }^{b}A_{\mu }^{c})+{{\textstyle \frac{1}{2}}}F_{a}^{\nu
\mu }f_{bc}^{a}\rule{0em}{3ex}\eta ^{b}F_{\nu \mu }^{c}.  \label{eq:e_5_73}
\end{eqnarray}

The last term in above relation vanishes, i.\,e.%
\begin{equation*}
F_{a}^{\nu \mu }f_{bc}^{a}\rule{0em}{3ex}\eta ^{b}F_{\nu \mu }^{c}=0, 
\end{equation*}%
since%
\begin{eqnarray*}
F_{a}^{\nu \mu }f_{bc}^{a}\rule{0em}{3ex}\eta ^{b}F_{\nu \mu }^{c} &=&{{%
\textstyle \frac{1}{2}}}k_{am}\sigma ^{\nu \alpha }\sigma ^{\mu \beta
}F_{\alpha \beta }^{m}F_{\nu \mu }^{b}f_{bc}^{a}\rule{0em}{3ex}\eta ^{c} \\
&=&{{\textstyle \frac{1}{2}}}\sigma ^{\nu \alpha }\sigma ^{\mu \beta
}F_{\alpha \beta }^{m}F_{\nu \mu }^{b}f_{mbc}\eta ^{c}=0,
\end{eqnarray*}%
while%
\begin{equation*}
\begin{array}{cc}
f_{mbc}=k_{am}f_{bc}^{a}, & f_{mbc}=-f_{bmc}.%
\end{array}%
\end{equation*}%
Therefore, we derive%
\begin{equation}
\delta a_{1}-\gamma ({{\textstyle \frac{1}{2}}}F_{a}^{\nu \mu
}f_{bc}^{a}A_{\nu }^{b}A_{\mu }^{c})=\partial _{\nu }(-F_{a}^{\nu \mu
}f_{bc}^{a}\rule{0em}{3ex}\eta ^{b}A_{\mu }^{c}).  \label{eq:e_5_75}
\end{equation}%
It shows%
\begin{equation}
\begin{array}{cc}
a_{0}=-{{\textstyle \frac{1}{2}}}F_{a}^{\nu \mu }f_{bc}^{a}A_{\nu
}^{b}A_{\mu }^{c}, & \overset{(0)}{{j}^{\mu }}=-F_{a}^{\nu \mu }f_{bc}^{a}%
\rule{0em}{3ex}\eta ^{b}A_{\mu }^{c}.%
\end{array}
\label{eq:e_5_76}
\end{equation}

The results for the first-order deformation are summarized as follows:%
\begin{equation}
a=-{{\textstyle \frac{1}{2}}}F_{a}^{\nu \mu }f_{bc}^{a}A_{\nu }^{b}A_{\mu
}^{c}-A_{a}^{\ast \mu }f_{bc}^{a}\rule{0em}{3ex}\eta ^{b}A_{\mu }^{c}+{{%
\textstyle \frac{1}{2}}}\eta _{a}^{\ast }f_{bc}^{a}\rule{0em}{3ex}\eta ^{b}%
\rule{0em}{3ex}\eta ^{c}.  \label{eq:e_5_77_1}
\end{equation}%
Finally, we derive%
\begin{equation}
S_{1}=\int d^{D}x\,\left( -{{\textstyle \frac{1}{2}}}F_{a}^{\nu \mu
}f_{bc}^{a}A_{\nu }^{b}A_{\mu }^{c}-A_{a}^{\ast \mu }f_{bc}^{a}\rule%
{0em}{3ex}\eta ^{b}A_{\mu }^{c}+{{\textstyle \frac{1}{2}}}\eta _{a}^{\ast
}f_{bc}^{a}\rule{0em}{3ex}\eta ^{b}\rule{0em}{3ex}\eta ^{c}\right) .
\label{eq:e_5_78}
\end{equation}%
The first-order deformations of the solution ($S_{1}$) of the master equation were determined for the action (\ref{eq:e_5_44}). It is seen that  gauge
generators are field independent, and are reduced to a sum of terms with antighost
numbers from zero to two.

\subsection{Higher-order Deformations}
\label{sec5_2}

We now consider the higher-order deformations of the master equation for the action 
(\ref{eq:e_5_31}). The second-order deformation ($\lambda ^{2}$) of the master
equation are determined from the solution $(S_{1},S_{1})+2sS_{2}=0$. Let us assume that 
\begin{equation}
S_{2}=\int d^{D}x\,b,
\end{equation}%
that takes the local form 
\begin{equation}
\Delta +2sb=\partial _{\mu }m^{\mu }.  \label{eq:e_6_1}
\end{equation}%
Using the equation (\ref{eq:e_5_77_1}) from Section \ref{sec5_1}, we calculate
$(S_{1},S_{1})$:
\begin{eqnarray}
(S_{1},S_{1}) \equiv \int d^{D}x\,\Delta &=& \left( \int d^{D}x\,a,\int
d^{D}y\,a\right)  \notag \\
&=&\int d^{D}xd^{D}y\,\left( \,a(x),\,a(y)\right) ,  \notag
\end{eqnarray}%
while employing the following relations%
\begin{equation}
\left( \eta ^{a}(x),\eta _{b}^{\ast }(y)\right) =\left( \eta _{b}^{\ast
}(y),\eta ^{a}(x)\right) =-\delta _{b}^{a}\delta ^{D}(x-y),
\end{equation}%
\begin{equation}
\left( A_{\mu }^{a}(x),A_{b}^{\ast \nu }(y)\right) =\left( A_{b}^{\ast \nu
}(y),A_{\mu }^{a}(x)\right) =-\delta _{b}^{a}\delta _{\mu }^{\nu }\delta
^{D}(x-y),
\end{equation}%
and the definitions 
\begin{equation}
\sigma ^{\alpha \rho }\sigma ^{\beta \lambda }k_{mg}f_{\rho \lambda
}^{g}\equiv \sigma ^{\alpha \rho }\sigma ^{\beta \lambda }k_{mg}(\partial
_{\rho }A_{\lambda }^{g}-\partial _{\lambda }A_{\rho }^{g}),
\end{equation}%
\begin{equation}
\int d^{D}x\,\delta ^{D}(x-y)f(x)\equiv f(y).
\end{equation}%
They lead to the following expression $\Delta $: 
\begin{eqnarray*}
\,\Delta &=&{-}f_{em}^{a}\rule{0em}{3ex}f_{np}^{e}\rule{0em}{3ex}\eta
_{a}^{\ast }\eta ^{m}\rule{0em}{3ex}\eta ^{n}\rule{0em}{3ex}\eta
^{p}-(f_{em}^{a}\rule{0em}{3ex}f_{np}^{e}+f_{en}^{a}\rule{0em}{3ex}%
f_{pm}^{e}+f_{ep}^{a}\rule{0em}{3ex}f_{mn}^{e})A_{a}^{\ast \mu }\rule%
{0em}{3ex}\eta ^{m}\rule{0em}{3ex}\eta ^{n}A_{\mu }^{p} \\
&&+(f_{en}^{a}\rule{0em}{3ex}f_{pm}^{e}-f_{em}^{a}\rule{0em}{3ex}%
f_{pn}^{e})F_{a}^{\alpha \beta }\rule{0em}{3ex}A_{\alpha }^{m}A_{\beta
}^{n}\eta ^{p}+f_{bc}^{a}\rule{0em}{3ex}k_{ma}f_{np}^{m}\sigma ^{\alpha \rho
}\sigma ^{\beta \mu }(\partial _{\rho }\eta ^{b}\rule{0em}{3ex})A_{\mu
}^{c}A_{\alpha }^{n}A_{\beta }^{p} \\
&&+f_{bc}^{a}\rule{0em}{3ex}k_{ma}f_{np}^{m}\sigma ^{\alpha \rho }\sigma
^{\beta \mu }\eta ^{b}\rule{0em}{3ex}(\partial _{\rho }A_{\mu
}^{c})A_{\alpha }^{n}A_{\beta }^{p}-f_{bc}^{a}\rule{0em}{3ex}%
k_{ma}f_{np}^{m}\sigma ^{\alpha \mu }\sigma ^{\beta \lambda }(\partial
_{\lambda }\eta ^{b}\rule{0em}{3ex})A_{\mu }^{c}A_{\alpha }^{n}A_{\beta }^{p}
\\
&&-f_{bc}^{a}\rule{0em}{3ex}k_{ma}f_{np}^{m}\sigma ^{\alpha \mu }\sigma
^{\beta \lambda }\eta ^{b}\rule{0em}{3ex}(\partial _{\rho }A_{\mu
}^{c})A_{\alpha }^{n}A_{\beta }^{p}),
\end{eqnarray*}%
that is reduced to 
\begin{eqnarray*}
\Delta &=&-{{\textstyle \frac{1}{3!}}}f_{e[m}^{a}f_{np]}^{e}\eta _{a}^{\ast
}\eta ^{m}\eta ^{n}\eta ^{p}-f_{e[m}^{a}f_{np]}^{e}A_{a}^{\ast \mu }\eta
^{m}\eta ^{n}A_{\mu }^{p} \\
&&-f_{e[m}^{a}f_{np]}^{e}F_{a}^{\alpha \beta }A_{\alpha }^{m}A_{\beta
}^{n}\eta ^{p}+2f_{bc}^{a}k_{ma}f_{np}^{m}\sigma ^{\alpha \rho }\sigma
^{\beta \mu }(\partial _{\rho }\eta ^{b})A_{\mu }^{c}A_{\alpha }^{n}A_{\beta
}^{p}.
\end{eqnarray*}

We then decompose $\Delta $ into the following terms,%
\begin{equation}
\Delta =\Delta _{0}+\Delta _{1}+\Delta _{2},
\end{equation}%
namely,%
\begin{equation}
\Delta _{0}\equiv -f_{e[m}^{a}f_{np]}^{e}F_{a}^{\alpha \beta }A_{\alpha
}^{m}A_{\beta }^{n}\eta ^{p}+2f_{bc}^{a}k_{ma}f_{np}^{m}\sigma ^{\alpha \rho
}\sigma ^{\beta \mu }(\partial _{\rho }\eta ^{b})A_{\mu }^{c}A_{\alpha
}^{n}A_{\beta }^{p}.  \label{eq:e_6_4}
\end{equation}%
\begin{equation}
\Delta _{1}\equiv -f_{e[m}^{a}f_{np]}^{e}A_{a}^{\ast \mu }\eta ^{m}\eta
^{n}A_{\mu }^{p},  \label{eq:e_6_3}
\end{equation}%
\begin{equation}
\Delta _{2}\equiv -{{\textstyle \frac{1}{3!}}}f_{e[m}^{a}f_{np]}^{e}\eta
_{a}^{\ast }\eta ^{m}\eta ^{n}\eta ^{p},  \label{eq:e_6_2}
\end{equation}%
We also define 
\begin{equation}
b\equiv b_{0}+b_{1}+b_{2}.
\end{equation}

From (\ref{eq:e_6_1}), it follows a set of equations%
\begin{eqnarray}
\Delta _{2}+2\gamma b_{2} &=&\partial _{\mu }\overset{(2)}{{m}^{\mu }},
\label{eq:e_6_5} \\
\Delta _{1}+2\delta b_{2}+2\gamma b_{1} &=&\partial _{\mu }\overset{(1)}{{m}%
^{\mu }},  \label{eq:e_6_6} \\
\Delta _{0}+2\delta b_{1}+2\gamma b_{0} &=&\partial _{\mu }\overset{(0)}{{m}%
^{\mu }}.  \label{eq:e_6_7}
\end{eqnarray}%
Equations (\ref{eq:e_6_2}) and (\ref{eq:e_6_5}) imply
\begin{equation}
\begin{array}{cc}
\Delta _{2}=0, & b_{2}=0,%
\end{array}%
\end{equation}%
and%
\begin{equation}
f_{e[m}^{a}f_{np]}^{e}=0.  \label{eq:e_6_8}
\end{equation}%
The later expression is called the Jacobi identity. Similarly, we obtain%
\begin{equation}
\begin{array}{cc}
\Delta _{1}=0, & b_{1}=0.%
\end{array}%
\end{equation}%
So, the equation (\ref{eq:e_6_7}) remains to be solved:%
\begin{equation}
2f_{bc}^{a}k_{ma}f_{np}^{m}\sigma ^{\alpha \rho }\sigma ^{\beta \mu
}(\partial _{\rho }\eta ^{b})A_{\mu }^{c}A_{\alpha }^{n}A_{\beta
}^{p}+2\gamma b_{0}=\partial _{\mu }\overset{(0)}{{m}^{\mu }}.
\end{equation}%
We solve it by substituting the exterior longitudinal differential $\gamma $
 of potentials $A_{\mu }^{a}$ ($\gamma A_{\mu }^{a}=\partial _{\mu }\eta
^{\alpha }$):%
\begin{eqnarray}
2f_{bc}^{a}k_{ma}f_{np}^{m}\sigma ^{\alpha \rho }\sigma ^{\beta \mu
}(\partial _{\rho }\eta ^{b})A_{\mu }^{c}A_{\alpha }^{n}A_{\beta }^{p}
=\gamma \left( -{{\textstyle \frac{1}{2}}}f_{bc}^{a}k_{am}f_{np}^{m}\sigma
^{\alpha \rho }\sigma ^{\beta \mu }A_{\rho }^{b}A_{\mu }^{c}A_{\alpha
}^{n}A_{\beta }^{p}\right) .  \notag
\end{eqnarray}%
Accordingly, we derive 
\begin{equation}
b_{0}=-{{\textstyle \frac{1}{4}}}f_{bc}^{a}k_{am}f_{np}^{m}\sigma ^{\alpha
\rho }\sigma ^{\beta \mu }A_{\rho }^{b}A_{\mu }^{c}A_{\alpha }^{n}A_{\beta
}^{p}.  \notag
\end{equation}%
Hence, the second-order deformations becomes%
\begin{equation}
S_{2}=\int d^{D}x\,\left( -{{\textstyle \frac{1}{4}}}%
f_{bc}^{a}k_{am}f_{np}^{m}\sigma ^{\alpha \rho }\sigma ^{\beta \mu }A_{\rho
}^{b}A_{\mu }^{c}A_{\alpha }^{n}A_{\beta }^{p}\right) .
\end{equation}

The Jacobi identity (\ref{eq:e_6_8}) obviously implies %
\begin{equation*}
(S_{1},S_{2})=0\rightarrow S_{3}=0. 
\end{equation*}%
Similarly, all deformations with orders higher than the second-order completely vanish:%
\begin{equation*}
S_{k}=0,\ \forall k\geqslant 3. 
\end{equation*}%
As a result, the solution to the deformations becomes ${S}={S}_{0}+\lambda {S}_{1}+\lambda ^{2}{S}_{2}$, that corresponds to the following Yang-Mills theory:%
\begin{eqnarray}
{S} &=&\int d^{D}x\left( -{{\textstyle\frac{1}{4}}}F_{\mu \nu
}^{a}F_{a}^{\mu \nu }+A_{a}^{\ast \mu }\partial _{\mu }\eta ^{a}\right) 
\notag \\
&&+\lambda \int d^{D}x\,\left( -{{\textstyle\frac{1}{2}}}F_{a}^{\nu \mu
}f_{bc}^{a}A_{\nu }^{b}A_{\mu }^{c}-A_{a}^{\ast \mu }f_{bc}^{a}\rule%
{0em}{3ex}\eta ^{b}A_{\mu }^{c}+{{\textstyle\frac{1}{2}}}\eta _{a}^{\ast
}f_{bc}^{a}\rule{0em}{3ex}\eta ^{b}\rule{0em}{3ex}\eta ^{c}\right)  \notag \\
&&+\lambda ^{2}\int d^{D}x\,\left( -{{\textstyle\frac{1}{4}}}%
f_{bc}^{a}k_{am}f_{np}^{m}\sigma ^{\alpha \rho }\sigma ^{\beta \mu }A_{\rho
}^{b}A_{\mu }^{c}A_{\alpha }^{n}A_{\beta }^{p}\right) .  \label{eq:e_6_9}
\end{eqnarray}%
We have determined the Yang-Mills theory from the first- and
second-order deformations of the master equation. The solutions of the master equation, which
entirely include the gauge structures, are decomposed into terms
with the antighost numbers from zero to two. In other words, the
part with the antighost number equal to zero represents the Lagrangian action,
while the antighost number one is proportional to the gauge generators. The
terms with higher antighost numbers provide the reducibility functions, where
 the on-shell relations become linear components in the ghosts for
ghosts. It is shown that all functions with order higher than second vanish
in this model.

\subsection{Interacting theory}

Let us consider the equation (\ref{eq:e_6_9}) and identify the entire gauge
structure of the Lagrangian model that describes all consistent
interactions in the $D$-dimensional free Yang-Mills theory.

The antighost number zero of (\ref{eq:e_6_9}) shall provide the Lagrangian
action of the interacting theory:%
\begin{eqnarray}
{S_{0}^{\mathrm{L}}}[A_{\mu }^{a}] &=&\int d^{D}x\left( -{{\textstyle\frac{1%
}{4}}}F_{\mu \nu }^{a}F_{a}^{\mu \nu }\right)  \notag \\
&&+\lambda \int d^{D}x\,\left( -{{\textstyle\frac{1}{2}}}F_{a}^{\nu \mu
}f_{bc}^{a}A_{\nu }^{b}A_{\mu }^{c}\right)  \notag \\
&&+\lambda ^{2}\int d^{D}x\,\left( -{{\textstyle\frac{1}{4}}}%
f_{bc}^{a}k_{am}f_{np}^{m}\sigma ^{\alpha \rho }\sigma ^{\beta \mu }A_{\rho
}^{b}A_{\mu }^{c}A_{\alpha }^{n}A_{\beta }^{p}\right) .  \label{eq:e_6_10}
\end{eqnarray}%
Accordingly, the Yang-Mills theory is characterized by the following non-abelian action:%
\begin{equation}
{S_{0}^{\mathrm{L}}}[A_{\mu }^{a}]=\int d^{D}x\left( -{{\textstyle\frac{1}{4}%
}}\mathcal{F}_{\mu \nu }^{a}\mathcal{F}_{a}^{\mu \nu }\right) ,
\end{equation}%
where the non-abelian field strengths $\mathcal{F}_{\mu \nu }^{a}$ is defined by%
\begin{equation}
\mathcal{F}_{\mu \nu }^{a}=F_{\mu \nu }^{a}+\lambda f_{bc}^{a}A_{\mu }^{b}A_{\nu
}^{c},
\end{equation}%
and $f_{bc}^{a}$ is the gauge-invariant that provides the gauge symmetry of the
Yang-Mills theory as follows%
\begin{equation}
\bar{\delta}_{\varepsilon }A_{\mu }^{a}=\partial _{\mu }\varepsilon
^{a}-\lambda f_{bc}^{a}\varepsilon ^{b}A_{\mu }^{c}\equiv D_{\mu
}\varepsilon ^{a}.  \label{eq:e_6_12}
\end{equation}%
So, the commutator among the deformed gauge
transformations becomes:%
\begin{equation}
\lbrack \bar{\delta}_{\varepsilon _{1}},\bar{\delta}_{\varepsilon
_{2}}]A_{\mu }^{a}=\bar{\delta}_{\varepsilon }A_{\mu }^{a}.
\label{eq:e_6_11}
\end{equation}%
The gauge symmetry remains abelian to order $\lambda $, and satisfies the equation of motion%
\begin{equation}
D^{\mu }\mathcal{F}_{\mu \nu }^{a}=0.
\end{equation}%
The invariance of the action under the gauge transformations (\ref{eq:e_6_12}%
) is also obtained by the Noether identities%
\begin{equation}
D^{\mu }\left( \frac{{\delta \mathcal{L}}_{0}}{{\delta }A_{\mu }^{a}}\right)
\equiv D^{\mu }D^{\nu }\mathcal{F}_{\mu \nu }^{a}=0.
\end{equation}%
The antighost number one of the deformation of the master equation allows 
to identify the gauge transformations (\ref{eq:e_6_12}) of the action (\ref%
{eq:e_6_10}) by substituting the ghost $\eta ^{a}$\ with gauge
parameter $\varepsilon ^{a}$. The antighost number two in (\ref{eq:e_6_9})
reads the complete gauge structure of the so-called \textit{interacting
theory} that determines the commutator (\ref{eq:e_6_11}) among the deformed
gauge transformations.

\section{Conclusion}

\label{sec6}

In this paper, we reviewed deformed gauge transformations 
in the framework of the BRST-antifield formalism characterized by the
antibracket that acts similar to the Poisson
bracket in the Hamiltonian formalism. We provided the BRST cohomology of the
consistent interactions through several order deformations of the master equation. The
BRST-antifield formalism in the cohomological space 
provides the generalized framework of consistent interactions among
fields with a gauge freedom by any types of invariant action.
We see that higher order deformations could be neglected due to non local
interactions and their obstruction of consistent local couplings, which are associated
with the anomalous gauge quantization. We demonstrated its functions by applying the BRST-antifield formalism to 
the $D$-dimensional, free Yang-Mills theory. 
All deformations of the master equation for the massless Yang-Mills model were calculated
 by using the cohomological groups $H_{I}(s|d)$, $I=0,\ldots ,2$, of the BRST differential. 
The first-order deformation is provided by the cohomological group $H_{1}(s|d)$, whereas the second-order deformation given by
the cohomological group $H_{2}(s|d)$ obstructs all higher-order deformations. The 
results show that the deformations can be synthesized by the
conception that all orders higher than two are trivial, while  
gauge generators are imposed to be field independent, $H_{I}(s|d)=0$, $I>2$. 
The deformations stopped at the second-order
of the coupling constants characterize the consistent
interactions, which maintain the equation of motion, and 
provide the entire gauge structure of the interacting
Yang-Mills theory.

 {\color{myaqua}

 \vskip 6mm

 \noindent\Large\bf Acknowledgments}

 \vskip 3mm

{ \fontfamily{times}\selectfont
 \noindent
 The author thanks the editor and the referee for their comments. 
 Research of A. Danehkar is funded by the EU contract MRTN-CT-2004-005104. This support is greatly appreciated.}


\newpage 

\sectionn{Appendix}

\subsectionn{Antibracket Structure}

\label{app_1}

For a function $X(\psi )$ in a generic space, commutative or
anticommutative, we state:
\begin{equation}
{\frac{{\partial _{l}X}}{{\partial \psi }}=\frac{{\vec{\partial}}}{{\partial
\psi }}X,}~~~{\frac{{\partial _{r}X}}{{\partial \psi }}=X\frac{{%
\mathord{\buildrel{\lower3pt\hbox{$\scriptscriptstyle\leftarrow$}} \over
\partial }}}{{\partial \psi }}.}  \label{eq:e_a_1}
\end{equation}%
The \textit{left derivative }${\partial _{l}}$ is an ordinary derivative
(left to right). The \textit{right derivative} ${\partial _{r}}$ is the
derivative action from right to left.

For any $X(\psi )$ in a generic space, we get
\begin{equation}
\frac{{\partial _{l}X}}{{\partial \psi }}=(-1)^{\varepsilon _{\psi
}(\varepsilon _{X}+1)}\frac{{\partial _{r}X}}{{\partial \psi }}.
\label{eq:e_a_3}
\end{equation}
Considering Eqs. (\ref{eq:e_4_6}) and (\ref{eq:e_a_3}), it follows that
\begin{equation}
(X,Y)=-(-1)^{(\varepsilon _{X}+1)(\varepsilon _{Y}+1)}(Y,X).  \notag
\end{equation}%
Assuming $X=Y$, one can find
\begin{equation}
\frac{{\partial _{r}X}}{{\partial \Phi ^{A}}}\frac{{\partial _{l}X}}{{%
\partial \Phi _{A}^{\ast }}}=(-1)^{(\varepsilon _{X}+1)(\varepsilon _{X}+1)}%
\frac{{\partial _{r}X}}{{\partial \Phi _{A}^{\ast }}}\frac{{\partial _{l}X}}{%
{\partial \Phi ^{A}}}.  \label{eq:e_a_6}
\end{equation}%
For bosonic (commutative) and fermionic (anticommutative) variables, we have
\begin{equation}
(X,X)=\left\{ {%
\begin{array}{cc}
{\displaystyle2\frac{{\partial _{r}X}}{{\partial \Phi ^{A}}}\frac{{\partial
_{l}X}}{{\partial \Phi _{A}^{\ast }}}} & {X\mathrm{~is~commutative,}} \\
0 & {X\mathrm{~is~anticommutative}\mathrm{.}}%
\end{array}%
}\right.  \label{eq:e_a_7}
\end{equation}%
For any $X$, we have
\begin{equation}
\begin{array}{cc}
{((X,X),X)=0,} & {\forall X.}%
\end{array}
\label{eq:e_a_8}
\end{equation}

Furthermore, the antibracket has the following properties:
\begin{equation}
(X,YZ)=(X,Y)Z+(-1)^{\varepsilon _{Y}\varepsilon _{Z}}(X,Z)Y,
\label{eq:e_4_8}
\end{equation}%
\begin{equation}
(XY,Z)=X(Y,Z)+(-1)^{\varepsilon _{X}\varepsilon _{Y}}Y(X,Z),
\label{eq:e_4_9}
\end{equation}%
\begin{eqnarray}
((X,Y),Z)+(-1)^{(\varepsilon _{X}+1)(\varepsilon _{Y}+\varepsilon
_{Z})}((Y,Z),X) &&  \notag \\
+(-1)^{(\varepsilon _{Z}+1)(\varepsilon _{X}+\varepsilon _{Y})}((Z,X),Y)
&=&0.  \label{eq:e_4_10}
\end{eqnarray}

\end{document}